\documentclass[onecolumn]{aastex6}

\newcounter{magicrownumbers}

\slugcomment{Accepted for publication in The Astronomical Journal}
\shorttitle{FUV Spectroscopy of ER UMa}
\shortauthors{Guzman et al.}
\watermark{ApJ Accepted Version}
\setwatermarkfontsize{0.9in} 

\begin{document}

\title{{\bf  
Fuse and IUE Spectroscopy of the Prototype Dwarf Nova ER Ursa Majoris During Quiescence      
} }

\author{Giannina Guzman\altaffilmark{1}} 
\author{Edward M. Sion\altaffilmark{1}} 
\author{Patrick Godon\altaffilmark{1,2}}

%\and

\email{gguzman2@villanova.edu}
\email{edward.sion@villanova.edu} 
\email{patrick.godon@villanova.edu}
 
\altaffiltext{1}{Department of Astrophysics \& Planetary Science, 
Villanova University, Villanova, PA 19085, USA}
\altaffiltext{2}{Henry A. Rowland Department of Physics \& Astronomy,
The Johns Hopkins University, Baltimore, MD 21218, USA}

\begin{abstract}
ER Ursae Majoris is the prototype for a subset of SU UMa-type dwarf novae 
characterized by short cycle times between outburst, high outburst frequency, 
and ``negative'' superhumps. It suffers superoutbursts every 43 days, 
lasting 20 days, normal outbursts every 4 days and has an outburst 
amplitude of 3 magnitudes.  
We have carried out a far ultraviolet (FUV) spectral analysis of 
ER UMa in quiescence, by fitting {\it Far Ultraviolet Spectroscopic 
Explorer (FUSE)} and {\it International Ultraviolet Explorer (IUE)} 
spectra with model accretion disks and high gravity photosphere models. 
Using the Gaia parallax distance and an orbital inclination of $50^{\circ}$, we find that during the brief 
quiescence of only four days, the accretion rate is   
$7.3 \times 10^{-11}M_{\odot}$/yr,  
with the ER UMa white dwarf contributing 55\% 
of the FUV flux and the accretion disk contributing the remaining 
45\% of the flux.  The white dwarf in ER UMa 
is markedly hotter (32,000~K) than the other white dwarfs in dwarf novae 
below the CV period gap which have typical temperatures $\sim$15,000~K. For a higher inclinations of 60 to 75 degrees, the accretion rates that we derive are roughly an order of magnitude higher $1 - 3 \times 10^{-10}M_{\odot}$/yr.  
\end{abstract}

\keywords{
--- dwarf novae, cataclysmic variables  
--- stars: white dwarfs  
--- stars: individual (ER UMa)  
}

\section{{\bf Introduction}} \label{sec:intro}

ER Ursae Majoris is the prototype system of a subclass of SU UMa type dwarf 
novae. Like SU UMa systems they exhibit two types of outbursts: normal and 
superoutbursts. These superoutbursts show superhumps that are ruled by tidal 
instabilities. In the canonical view of the CVs below the period gap, the 
mass-transfer rate from the secondary star is governed by the rate of angular 
momentum loss due to gravitational wave emission, which happens to be strongly 
correlated to the orbital period. Yet, there were some systems that challenged 
the traditional SU UMa classification due to their frequency of outbursts, 
superoutbursts and suspected unusually high mass-transfer rates for the dwarf 
novae below the CV period gap. Once ER UMa was discovered in the Palomar Sky 
Survey as an ultraviolet-excess object \citep{gre86}      , along with a handful 
of similar systems, it was clear that they belonged to a new subclass of SU UMa systems  
(e.g. Iida 1994; Kato \& Jayacura 1995; Kato et al. 1999). After the observation of dwarf nova 
outbursts in the system (Iida 1994), it was seen that ER UMa is characterized 
by an extremely high outburst frequency and short supercycles (Kato et al. 1999), 
suggesting a higher mass transfer rate than other SU UMa sytems. During a 
superoutburst, ER UMa was confirmed to have `negative' superhumps, a behavior 
observed since the 1990's. Negative superhumps are a dynamic behavior of the 
system in which the superhumps have a shorter period than the orbital period 
(Ohshima et al. 2018), and displayed retrograde precession 
(Harvey et al. 1995, 
Patterson et al. 1997, Patterson 1999, Skillman et al. 1999, Wood \& Burke 2007, 
Montgomery 2012). This is thought to be due to an eccentric tilted disk. 
Some theories suggest that ER UMas are simply an evolutionary stage of classical 
novae, through studies of the recently sub-classified ER UMa system: BK Lyncis 
(Patterson et al. 2012).

The published orbital and physical parameters of ER UMa are given in Table 1 along with the literature references. \citet{dub18} take an inclination of $45 \pm 10 ^{\circ}$ for ER UMa, but the observation of sharp absorption lines
in the optical may indicate that the system may have a low inclination \citep{szk96}. The correct orbital inclination of ER UMa is unknown.   
Szkody et al.(1996) cite the line widths, small equivalent widths and 
low radial velocity amplitude of ER UMa as indicating a low inclination. 
On the other hand, Thorstensen et al. (1997) found, using the double 
convolution method, that the separation of the Gaussian peaks is 1260 km/s 
and the radial velocity of the WD is 48 km/sec. 
If one assumes $1 M_{\odot}$ for the WD and $0.1 M_{\odot}$ for the 
secondary star and applies Kepler's 3rd Law, the corresponding 
inclination is $i \sim 50^{\circ}$. However, the uncertainties in this 
inclination value are large. First, the H$_\alpha$ radial velocity curve 
of the white dwarf obtained over four nights shows a large scatter. 
Thorstensen et al.(1997) caution that, in cases where there is an 
independent check on the gamma velocity and radial velocity semi-amplitde 
K of the WD, these two quantities rarely reflect the true dynamical 
motion of the WD. Furthermore, different emission lines may form in 
different parts of the accretion disk and thus it cannot be assumed 
that they manifest the true dynamical motion of the white dwarf. 
Therefore, in view of the lack of a truly reliable value for the inclination, 
we carry out here an analysis of ER UMa FUV spectra 
for a range of values of the inclination,
namely $i=18^{\circ},~41^{\circ},~65^{\circ}$, and $70^{\circ}$.

In section 2, we present the FUV spectroscopic observations. In section 3, 
we present the details of our accretion disk and high gravity photospheric 
models, and we describe our analysis and model 
fitting results. Finally in section 4, we summarize our conclusions.

\begin{deluxetable*}{ccl} 
\tablewidth{0pt}
\tablecaption{Orbital and Physical Parameters of ER UMa}
\tablehead{
Parameter   & Value            & Reference \\  
}  
\startdata
$P$          & 0.06366~d       & \citet{tho97}  \\
$d$          & 374~pc          & Gaia     \\
$i$          & $18-50^{\circ}$ & \citet{szk96} ; \citet{dub18}     \\        
$E(B-V)$     & 0.01            &    \\  
$M_{\rm wd}$ & $1.0\pm0.2M_{\odot}$ & this paper \\
$M_2$        & 0.10$M_{\odot}$ & \citet{dub18}     \\
$q$          & 0.100           & Ohshima et al.2014  \\  
\enddata
\end{deluxetable*}

\section{{\bf Far Ultraviolet Spectroscopic Observations}}

In 1995, ER UMa was observed with {\it IUE} and seven spectra were
obtained during its 43 days supercycle \citep{szk96}. 
The data show large flux changes through the cycle with corresponding large spectral 
lines changes. The {\it IUE} spectrum presented here was obtained  during
quiescence on April 17 and exhibits mainly emission lines.    
The data were collected through the LARGE aperture, LOW dispersion, 
short wavelength camera: SWP54455. No data were collected through 
the long wavelength camera, hence we do not have long wavelength coverage
of the quiescent state of ER UMa.
The {\it IUE} spectrum of
ER UMa in quiescence has a wavelength coverage of $\sim$1150-2000~\AA . 
The exposure time is 8,700~s, or about $\sim 1.6 $ orbital period. 
The {\it IUE} observation log for the SWP54455 data is in Table 2.  

On January 15, 2004, ER UMa was observed in quiescence with {\it FUSE} \citep{fro12}  
with the LWRS (30 arcsec) aperture for 15
consecutive {\it FUSE} orbits. However, only 10 exposures had
valid data. The {\it FUSE} spectrum presented here consists of the
co-added 10 good exposures, totalling 30,625~s of good exposure time. 
The total (raw) {\it FUSE} observation time was 13.5~hr, or about $\sim 8.8$
orbital period.  
The {\it FUSE} observation log for the 10 exposures is in Table 2.  

All spectral data were retrieved from the on-line 
MAST archive and were all processed and calibrated by the pipelines.
For the {\it FUSE} observations we used our suite of IRAF procedures, 
FORTRAN programs, and Linux shell scripts to post-process the
data from the 8 different channels into one final spectrum
\citep[see ][e.g. taking care of the ``worm'']{god12}, 
with a wavelength coverage 904-1188 \AA .   
Due to the hydrogen cut-off (Lyman series/jump) the spectrum
starts around 914 \AA .      

The {\it FUSE} and {\it IUE} spectra were de-reddened assuming E(B-V) = 0.01 
and using the extinction curve of \citet{fit07}. 

\begin{deluxetable*}{ccccc} 
\tablewidth{0pt}
\tablecaption{Observation Log}
\tablehead{
Telescope   & Data ID     & Date (UG)  & Time (UT) & Exp. Time \\ 
            &             & yyyy-mm-dd & hh:mm:ss &  (s)       \\  
} 
\startdata
{\it IUE}   & SWP54455    & 1995-04-17 & 22:04:26 & 8700       \\ 
{\it FUSE}  & D90514002   & 2004-01-15 & 08:56:38 & 806         \\ 
{\it FUSE}  & D90514003   & 2004-01-15 & 09:42:18 & 2249        \\ 
{\it FUSE}  & D90514004   & 2004-01-15 & 10:38:29 & 1169        \\ 
{\it FUSE}  & D90514005   & 2004-01-15 & 11:30:07 & 3615        \\ 
{\it FUSE}  & D90514006   & 2004-01-15 & 13:06:43 & 4119        \\ 
{\it FUSE}  & D90514007   & 2004-01-15 & 14:43:52 & 4204        \\ 
{\it FUSE}  & D90514008   & 2004-01-15 & 16:24:22 & 4178        \\ 
{\it FUSE}  & D90514009   & 2004-01-15 & 18:02:43 & 4274        \\ 
{\it FUSE}  & D90514011   & 2004-01-15 & 19:57:13 & 3161        \\ 
{\it FUSE}  & D90514012   & 2004-01-15 & 21:45:09 & 2910        \\ 
\enddata
\end{deluxetable*}

In Figs. 1 and 2, we display the dereddened FUSE and IUE spectra, 
respectively, the continuum fluxes levels in both spectra matched 
up in the wavelengths region where they  overlap between 1170 and 1180. 
Thus, the two spectra combined together cover a broader wavelength 
range than the {\it FUSE} or {\it IUE} spectrum alone. 
This wider wavelength coverage samples more of the SED of ER UMa 
which helps to achieve more accurate model fits. 
Figs. 1 and 2 show the strongest lines identified for both spectra. 

Table 3 highlights the characteristics of the identified lines in the FUSE spectrum intrinsic to CVs, and the lines due to the ISM  have 
been labeled as such \citep[e.g.][]{god12}. 
Some characteristic 
lines are not clearly identifiable in the spectral plot 
and therefore are not included in the table. 

All the sharp emission lines (H\,{\sc i} and He\,{\sc ii}) 
in the {\it FUSE} spectrum of ER UMa are all likely due to 
daylight reflected inside the telescope (more than 50\% of the observation took 
place during the ``daytime'' of the {\it FUSE} telescope, i.e. when it is not
in the shade of the Earth). The very broad emission lines of 
O\,{\sc vi} (1032 \& 1038 \AA ) and C\,{\sc iii} (1175 \AA )
are from the hottest component of the system located in the inner disk
and exhibit a strong Keplerian broadening.  
There are possibly additional broad emission lines (e.g. C\,{\sc iii} 977 \AA )
especially in the very short wavelength region ($\lambda < 950$ \AA ,
higher ionization species such as S\,{\sc vi}, N\,{\sc iv}) which might be 
broad enough to merge together to form the continuum
observed in the range $\sim 920-945$ \AA .  
The spectrum displays many sharp and shallow 
ISM absorption lines mostly from molecular hydrogen (rotational and vibrational
energy levels) and from some metals
such as  N\,{\sc i}, C\,{\sc ii}, Si{\sc ii}, Fe\,{\sc ii},
and Ar\,{\sc i}. 
The S\,{\sc iv}, Si\,{\sc iii} \& {\sc iv}
and C\,{\sc iii} absorption lines are from ER UMa as they are not
as sharp. Taking into account that 
the FUSE spectrum was obtained over a period of time of the order
of the several binary orbital periods, the fact that we do see
narrow absorption lines from the source (but not as narrow as the ISM lines)is an additional argument in favour of the low inclination.  

Table 4 shows the identifiable lines from 
the IUE spectrum. 
The very strong S\,{\sc iv} emission doublet is due to the high 
temperature of the hot component
in the system. C\,{\sc iv} is in emission too, and could be associated with a disk corona or the boundary layer. 
Finally, there is also a distinct blend of 
Al\,{\sc iii} lines, which, because of their apparent strength, could be due to a suprasolar abundance of Al III, perhaps accumulated from the thermonuclear ashes of  previous novae that are either being fed back to the 
white dwarf, or are brought to the surface of the white dwarf from below by a 
process such as radiative acceleration or forced (shear) convective mixing 
during outburst.

\clearpage 

\begin{figure}
\plotone{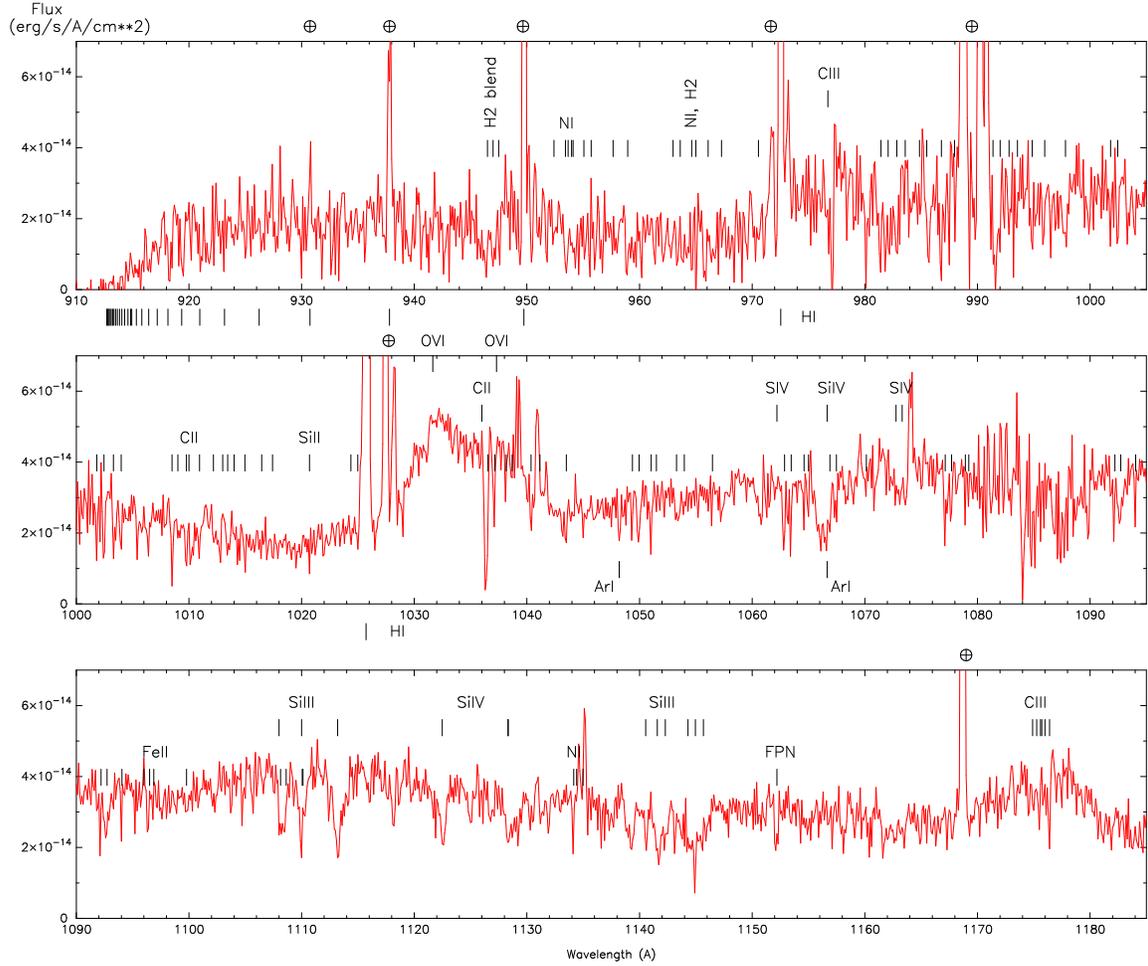} 
\caption{
{\it FUSE} spectrum of ER UMa (flux versus wavelength) obtained 
during quiescence. The sharp emission lines (H\,{\sc i} and He\,{\sc ii}) 
are all due to daylight
reflected inside the telescope, and only the broad emission lines 
of O\,{\sc vi} (1032 \& 1038 \AA ) and C\,{\sc iii} (1175 \AA )
are positively identified as originating from the source.  
ISM absorption lines have been marked with tick marks in the middle
of each panel and consist mostly of molecular hydrogen. Additional ISM
absorption lines of N\,{\sc i}, C\,{\sc ii}, Si{\sc ii}, Fe\,{\sc ii},
Ar\,{\sc i} have also been marked. The S\,{\sc iv}, Si\,{\sc iii} \& {\sc iv}
and C\,{\sc iii} absorption lines are from ER UMa. 
The air glow emission lines have been annoted above each panel with a
cross inside a circle, and correspond mainly to the Lyman hydrogen series
which have been annotated below each panel. 
FPN is a detector fixed pattern noise artifact.  
}
\end{figure} 

\clearpage 

\begin{deluxetable*}{cccc} 
\tablewidth{0pt}
\tablecaption{FUSE Identified Line Features}
\tablehead{
Feature  & Wavelength  & Emission/Absorption & Likely Source}
\startdata
H\,{\sc i}   & 913-9       & e       & daylight \\ 
H\,{\sc i}   & 923.2       & e       & daylight \\ 
H\,{\sc i}   & 926.2       & e       & daylight \\ 
H\,{\sc i}   & 930.7       & e       & daylight \\ 
H\,{\sc i}   & 937.8       & e       & daylight \\ 
H\,{\sc i}   & 949.7       & e       & daylight \\ 
H\,{\sc i}   & 972.5       & e       & daylight \\ 
H$_2$ blend & 946-948     & a       & ISM \\ 
N\,{\sc i}   & 953-955     & a       & ISM \\ 
N\,{\sc i}   & 954-956     & a       & ISM \\ 
H$_2$ and N\,{\sc i} & 962-971   & a       & ISM \\ 
C\,{\sc iii} & $\sim$976.8 & a       & ISM     \\ 
N\,{\sc iii} & $\sim$989.6 & e       & daylight \\   
He\,{\sc ii} & 991.8       & a       & daylight\\ 
C\,{\sc ii}  & 1010        & a       & ISM \\ 
Si\,{\sc ii} & 1021        & a       & ISM \\ 
O\,{\sc vi}  & 1031.9      & e broad & Stellar \\ 
C\,{\sc ii}  & 1036        & a       & ISM   \\ 
O\,{\sc vi}  & 1037.6      & e broad & Stellar \\ 
Ar\,{\sc i}  & 1048.5      & a       & ISM \\ 
Si\,{\sc iv} & 1066.       & a       & Stellar \\ 
Ar\,{\sc i}  &  1066.8     & a       & ISM \\ 
S\,{\sc iv}  & 1073.4      & a + e?  & Stellar \\ 
Si\,{\sc iii} & 1108-1113   & a       & Stellar \\ 
P\,{\sc v} ?    & 1117        & a       & Stellar \\ 
Si\,{\sc iv} & 1122.5+1128 & a       & Stellar \\ 
P\,{\sc v} ?    & 1128        & a       & Stellar\\ 
N\,{\sc i}   & 1135        & e       & daylight \\ 
Si\,{\sc iii} blend & 1140.7-1145.9 & a & Stellar \\ 
FPN      & 1152.6      &         & Fixed Pattern Noise \\ 
He\,{\sc ii} & 1168        & e       & daylight \\ 
C\,{\sc iii} & 1175        & a/e?    & Stellar \\ 
\enddata
\end{deluxetable*}

\clearpage

\begin{figure}
\plotone{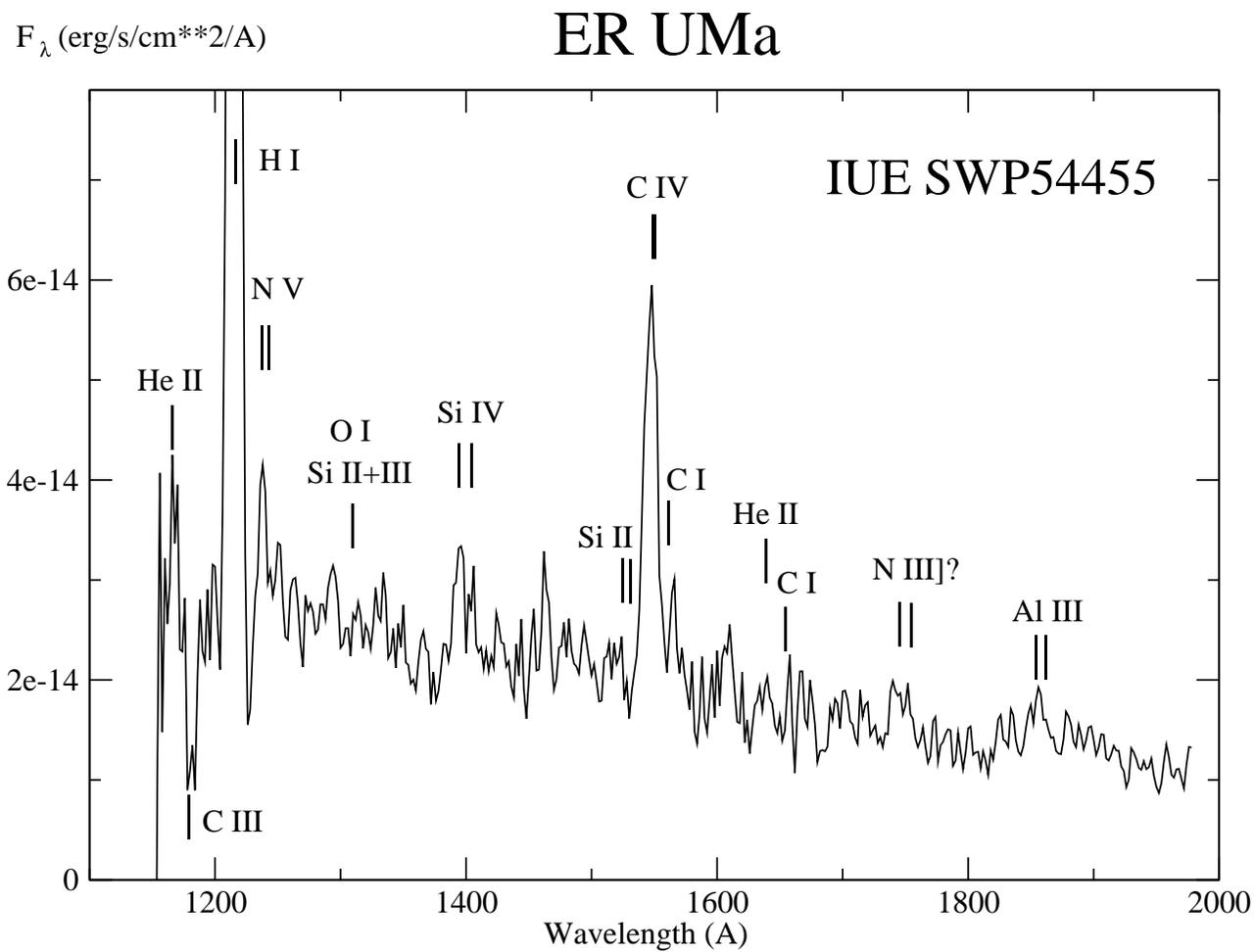} 
\caption{
IUE spectrum of ER UMa (Flux versus wavelength) obtained during dwarf nova 
quiescence with line identifications. 
}
\end{figure}

\clearpage

\begin{deluxetable*}{cccc} 
\tablewidth{0pt}
\tablecaption{IUE Identified Line Features}
\tablehead{
Feature  & Wavelength  & Emission/Absorption & Likely Source}
\startdata
He\,{\sc ii}  & 1168           & e   & stellar? \\  
C\,{\sc iii}   & 1175          & a   & stellar  \\	
H\,{\sc i}     & 1225          & e   & airglow  \\
N\,{\sc v}     & 1239-1243     & e/a & stellar  \\
O\,{\sc i} + S\,{\sc ii+iii} & $\sim$1300 & a   & stellar  \\	
Si\,{\sc iv}   & 1394, 1403    & e   & stellar  \\	
Si\,{\sc ii}   & 1527, 1533    & a   & stellar  \\ 
C\,{\sc iv}    & 1548-51       &	e   & stellar  \\
C\,{\sc i}     & 1561          & e/a & stellar  \\	
He\,{\sc ii}   & 1640          & e?  & stellar \\ 
C\,{\sc i}     & $\sim 1657$   & e/a?& stellar  \\	
Al\,{\sc iii}  & 1671          & e/a?& stellar \\
N\,{\sc iii]}  & 1747,1754     & e   & stellar \\ 	
Al\,{\sc iii}  & 1854, 1862    & e   & stellar \\
\enddata
\end{deluxetable*}

\clearpage

\section{{\bf Synthetic Spectral Modeling and Analysis.}}

The dereddened spectra were fitted with both disk and photosphere models in 
order to extract certain values for analysis such as white dwarf mass, 
accretion rate, and inclination angle. The model accretion disks were 
implemented from the optically thick, steady state, disk model grid for solar 
composition better known as the `standard disk model' (Wade and Hubeny 1998). 
For these models, the accretion disk's outermost radius $R_{out}$ is chosen 
in such a way that the effective temperature is around 10,000 K. 
Disk annuli beyond $R_{out}$ are neglected because they contain cooler zones 
with very little contribution to UV flux in the FUSE and IUE SWP 
spectral range.

The Wade \& Hubeny DISK models cover the following combination 
of inclination angle $i$, white dwarf mass $M_{\rm wd}$, and 
mass accretion rate $\dot{M}$:
$i$= 18, 41, 60, 75, \& 81 deg; 
$M_{\rm wd}= 0.35, 0.55, 0.80, 1.03, 1.21 M_{\odot}$, 
$Log(\dot{M})$ = -8.0,-8.5,-9.0,-9.5,-10.0,-10.5 $M_{\odot}$/yr. 
For the photosphere models, we used TLUSTY (Hubney 1988) and SYNSPEC   
(Hubney and Lanz 1995) to construct a solar composition, white dwarf, 
stellar photospheres grid. The temperatures range from 12,000 K to 60,000 K 
in 1,000 K to 5,000 K step sizes. 
The stellar surface gravity is set to   
agree with the white dwarf mass accretion disk models above. 
The projected stellar rotation rate is varied from 50 km/s to 500 km/s,
in steps of 50km/s. 

\subsection{Low Inclination Models} 

Our disk and photosphere modeling starts with the parameters listed 
in Table 1. Having the Gaia distance removes one free parameter.  
Optical spectroscopic observations \citep{szk96,tho97} 
reveal narrow line widths and low radial velocities 
in ER UMa ($K_1 \sim 50$km/s), which may be 
indicating the system orbital inclination may be low. 
Therefore, we first restricted the range of disk inclination angles in our disk modeling to be 18 degrees. The mass ratio q = 0.10 was taken from the study of the period of Stage A (positive) superhumps by Ohshima et al. (2014). We included a white dwarf component in the model fits described below, not only to assess its flux contribution in quiescence relative to an accretion disk but also because its inclusion should help improve the fits to the observed absorption lines that are unaccounted for in the disk model itself. 

With the Gaia distance $d = 374$~pc,  a disk inclination angle 
$i=18^{\circ}$, we fit the observed combined {\it FUSE + IUE} 
spectrum with a
white dwarf mass $M_{\rm wd} \approx 1.0 \pm 0.2 M_{\odot}$, with a temperature
of $T_{\rm wd} = 30,000 \pm 5000$~K, and a disk with a mass accretion
rate $\dot{M} = 10^{-10.5}-10^{-10} M_{\odot}$/yr.  
Fitting the {\it FUSE} and {\it IUE} spectra alone gives the same results. 

In Figs. 3 \& 4 we display the best-fit accretion disk + WD model to the 
combined {\it FUSE + IUE} spectrum of ER UMa. For convenience the {\it FUSE}
spectral range is shown in Fig.3 and the {\it IUE} spectral range 
is shown in Fig.4. This model has a white dwarf mass 
$M_{\rm wd}= 1.03 M_{\odot}$, a mass accretion rate 
$\dot{M} = 3.26 \times 10^{-11}M_{\odot}$/yr and a white dwarf surface 
temperature $T_{\rm wd}=32,000$~K. 
The white dwarf, with a radius of 5,611~km ($Log(g)=8.6377$), 
contributes 55\% of 
the FUV flux and the accretion disk contributes 45\%. 
To fit the absorption lines, the projected stellar rotational velocity
$V_{\rm rot} \sin{i}$ was set to 100~km/s.  

In the very short wavelengths of the {\it FUSE} range, Fig.3, 
the model has too little flux to fit the observed continuum 
flux level. Models with a higher WD temperature and/or higher mass 
accretion rate  provides an adequate flux level there, 
but the distance obtained is far too large and the model steep slope 
does not agree with the {\it IUE} slope of the spectrum. These models
also generate different absorption lines that were not observed.
These models had to be discarded.
Instead, it is more likely that the flux continuum in the very short wavelengths
of {\it FUSE} is due to broad emission lines of N\,{\sc iv} ($\sim$923 \AA ),
S\,{\sc vi} (933 \& 945 \AA ) merged together.

\clearpage

\begin{figure}
\plotone{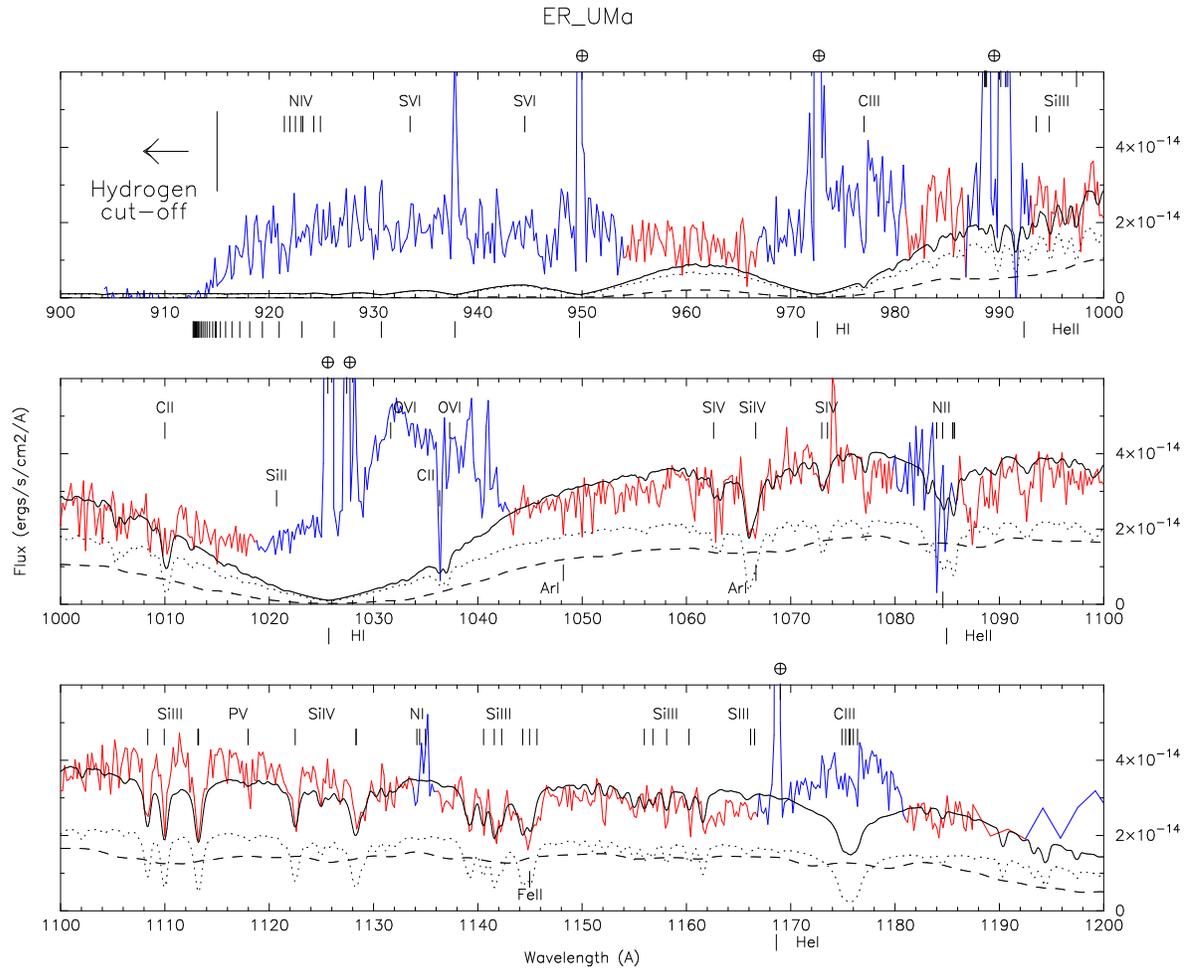}  
\caption{
A model fit (solid black line) to the de-reddened 
combined {\it FUSE + IUE} spectrum of ER UMa in quiescence (solid red line). 
The model is  a combined 32,000 K WD (dotted black line)
plus an accretion disk (dashed black line) with a mass accretion of 
$3.26 \times 10^{-11} M_{\odot}$/yr.  
The WD mass is $M_{\rm wd}= 1.03 M_{\odot}$, the inclination of the
disk is 18 degrees, and the distance has been set to 374~pc. 
The strong emission lines and regions affected by airglow have
been masked and are in blue. 
In order to match the absorption lines, the projected stellar rotational
velocity has been set to 100~km/s.  
}
\end{figure} 

\clearpage 

\begin{figure}
\vspace{-15.cm} 
\plotone{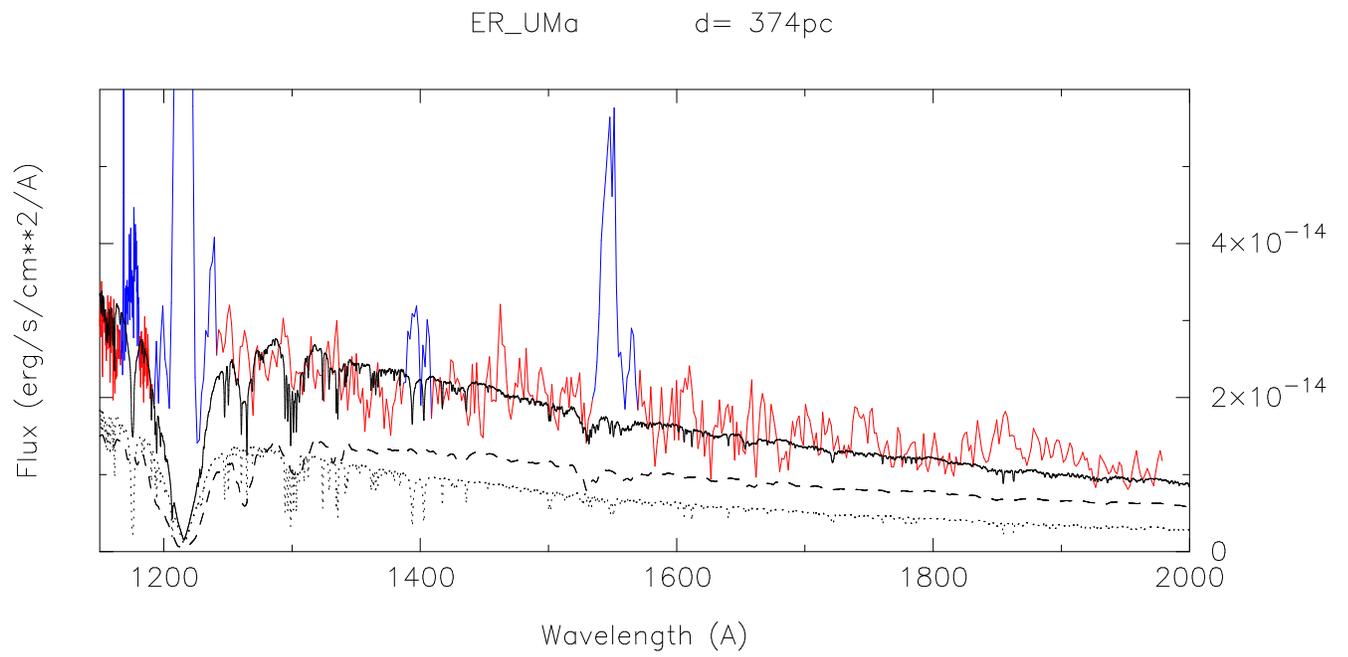}   
\caption{
Same as in Fig.3 displaying the {\it IUE} spectral range. 
}
\end{figure} 

\clearpage 

\subsection{High Inclination Models} 

In Fig. 5, we display the best-fit model accretion disk to the FUSE spectra 
of ER UMa. This model has a white dwarf mass of $M_{\rm wd}= 1.03 M_{\odot}$, 
an inclination angle of 75 degrees and yields an accretion rate of 
$\dot{M} = 3 \times 10^{-10}M_{\odot}$/yr. The scale-factor-derived distance 
(344 pc) is well within the error bars of the Gaia parallax distance. Next we 
sought the best-fit models to the combined FUSE plus IUE spectrum. 
This led to a best-fit disk model to the entire FUSE + IUE SWP 
spectrum displayed in Fig. 6 where again the white dwarf mass is 
$M_{\rm wd}= 1.03M_{\odot}$, but now the disk inclination is 60 degrees, 
the accretion rate is $\dot{M} = 1 \times 10^{-10}M_{\odot}$/yr and the 
scale-factor-derived distance is 370 pc. In Fig. 7, the combined  spectrum 
is best-fitted with a disk model having $M_{\rm wd}= 1.03 M_{\odot}$, disk 
inclination angle of 75 degrees and an accretion rate of 
$\dot{M} = 3 \times 10^{-10}M_{\odot}$/yr, corresponding to a scale factor-derived 
distance of 378 pc.
In both of the best fits to the combined spectra, the flux contribution 
of a white dwarf, although a minor contributor to the continuum flux, does 
help to improve the accuracy of the fit by accounting for the observed absorption lines.
We note also that the best fits to the combined (FUSE + IUE) spectrum 
give essentially the same accretion rate as that derived for the FUSE-only 
spectrum in Fig.5. 

\clearpage

\begin{figure}
\vspace{-05.cm} 
\plotone{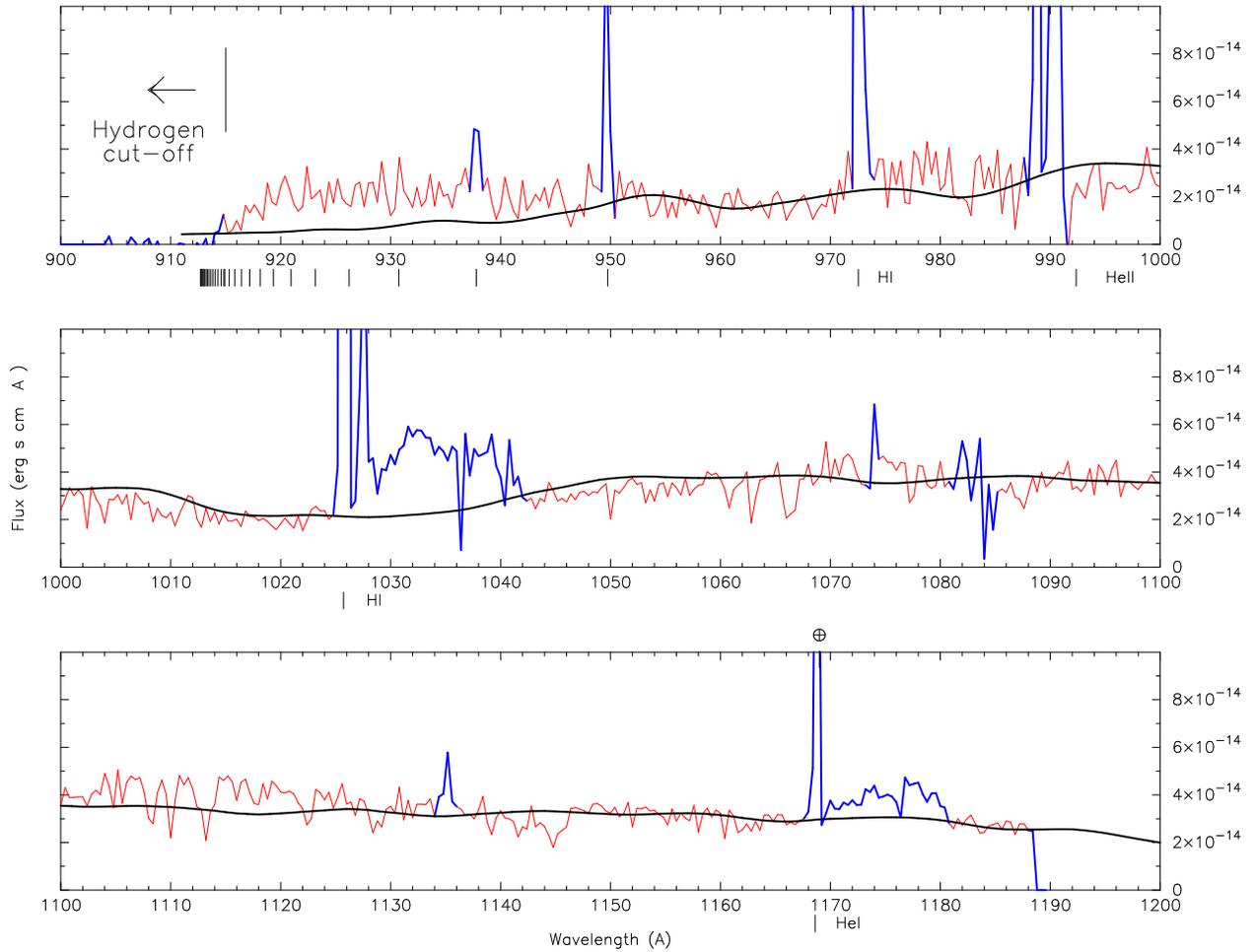} 
\caption{
The best-fit accretion disk model to the de-reddened FUSE spectrum (alone) 
of ER UMa in quiescence. The solid blue line is the disk model with 
$M_{\rm wd}= 1.03 M_{\odot}$, inclination angle of 75 degrees, and 
an accretion rate of $\dot{M} = 3.0 \times 10^{-10}M_{\odot}$/yr, 
giving a scaled distance of 340 pc.
}
\end{figure} 

\clearpage 

\begin{figure}
%\vspace{-05.cm} 
\plotone{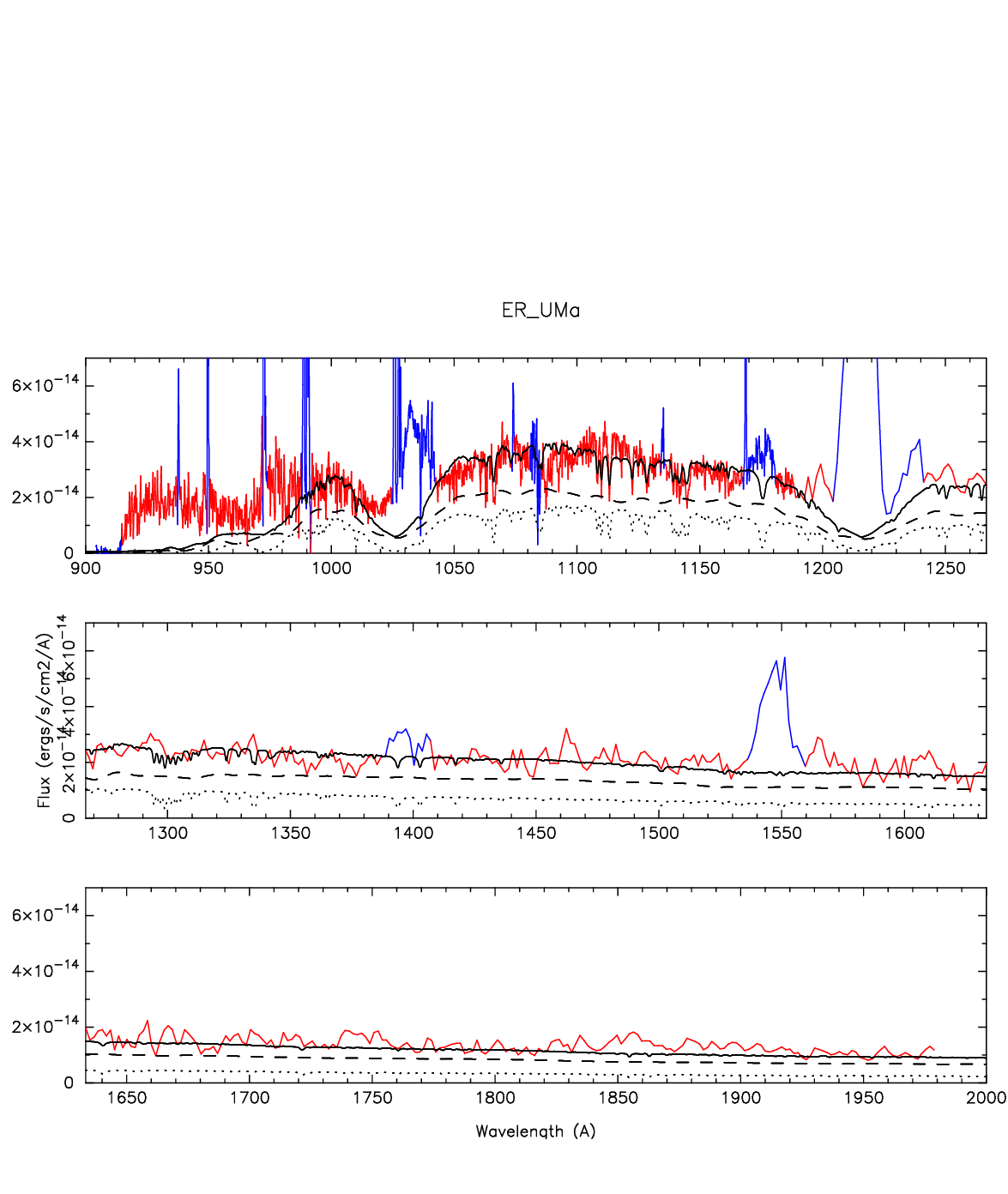} 
\caption{
The best-fit combined accretion disk plus white dwarf photosphere model 
to the de-reddened FUSE + IUE spectrum of ER UMa in quiescence. The solid 
line is the total model flux (disk plus white dwarf), the dashed line is the 
accretion disk flux, the dotted line is the flux contribution of the white 
dwarf. The disk parameters are 
$M_{\rm wd}= 1.03 M_{\odot}$, disk inclination angle of 60 degrees, and an 
accretion rate of $\dot{M} = 1\times 10^{-10}M_{\odot}$/yr. The white dwarf 
has Teff = 30,000K. The model distance is 370 pc
}
\end{figure} 

\clearpage

\begin{figure}
\plotone{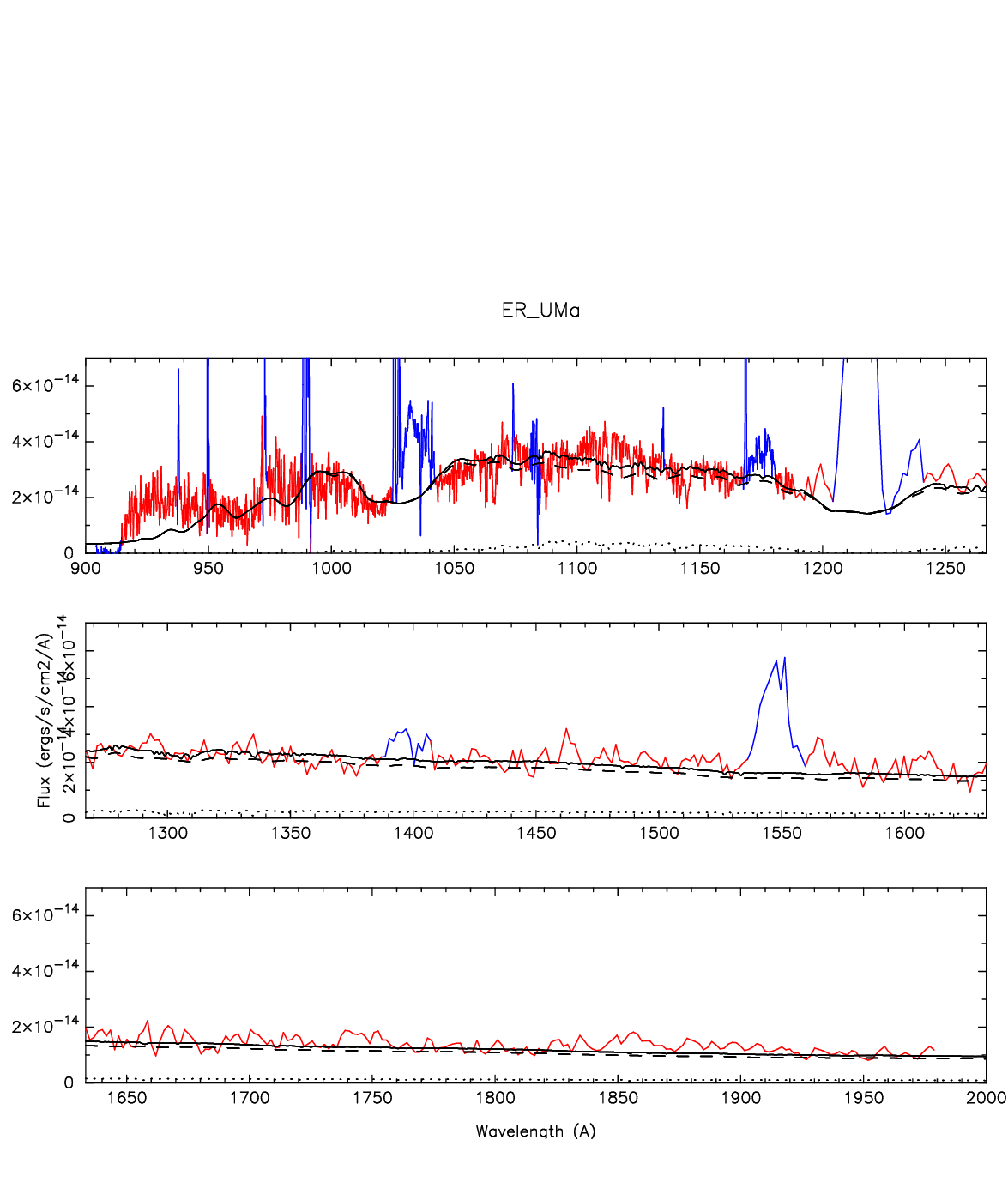} 
\caption{
The best-fit combined accretion disk plus white dwarf photosphere model 
to the de-reddened FUSE + IUE spectrum of  ER UMa in quiescence. The solid 
line is the total model flux (disk plus white dwarf), the dashed line is the 
accretion disk flux, the dotted line is the flux contribution of the white 
dwarf.  The disk parameters $M_{\rm wd}= 1.03 M_{\odot}$, disk inclination 
angle of 75 degrees, and an accretion rate of 
$\dot{M} = 3 \times 10^{-10}M_{\odot}$/yr. The white dwarf has Teff = 22,000K. 
The model distance is 378 pc.
}
\end{figure}

\clearpage

\begin{deluxetable*}{ccccccc} 
\tablewidth{0pt}
\tablecaption{Synthetic Spectra Model Fit Results}
\tablehead{
Model  & $M_{\rm wd}$ & $T_{\rm wd}$  & $\dot{M}$      & $i$ &  $d$ & Fig    \\ 
\#     & $M_{\odot}$  & 1000~K        & $M_{\odot}$/yr & deg &  pc  & \#       
}  
\startdata
  1   &  1.03     & 32       & $3.26 \times 10^{-11}$ & 18  & 374 & 3,4  \\  
  2   &  1.03     & ---      & $7.0 \times 10^{-11}$  & 41  & 381 & ---  \\  
  3   &  1.03     & 32       & $4.5 \times 10^{-11}$  & 41  & 367 & ---  \\  
  4   &  1.03     & 30       & $1.0 \times 10^{-10}$  & 60  & 370 &  6   \\  
  5   &  1.03     & ---      & $3.0 \times 10^{-10}$  & 75  & 344 &  5   \\  
  6   &  1.03     & 22       & $3.0 \times 10^{-10}$  & 75  & 378 &  7   \\  
\enddata
\end{deluxetable*}

\clearpage 

\section{{\bf Summary and Conclusion}} 

In Table 5, we tabulate the best-fit accretion disk models
for the two cases of high and low inclination. 
For an intermediate inclination of $i = 50^{\circ}$, 
we ran models with 
$i=41^{\circ}$  and $60^{\circ}$ which nicely bracket $50^{\circ}$. 
Interpolating between the $41^{\circ}$ disk model and the 
$60^{\circ}$ disk model, we estimate that the accretion rate 
of ER UMa in quiescence is $7.3 \times10^{-11} M_{\odot}$/yr
for a disk inclination angle of $50^{\circ}$, 
a $1.0 M_{\odot}$ WD and the Gaia distance of 374 pc.

With a distance of 374~pc, 
and for a WD mass range $M_{\rm wd} \approx 1.0 -1.2 M_{\odot}$, 
we find that during the brief quiescence of only four days, 
the accretion rate has dropped to  $10^{-10.5}-10^{-10} M_{\odot}$/yr, while the white dwarf has a temperature of $30,000 \pm 5000$~K. This accretion rate agrees with DN quiescent mass accretion rates, however, the white dwarf in ER UMa is much 
hotter ($\sim$30,000~K) than other white dwarfs in dwarf novae below the CV period gap which have typical temperatures $\sim$15,000K. 

If the white dwarf in ER UMa is heated by compressional 
heating alone, then the surface temperature during quiescence 
(Townsley \& Bildsten 2003) with an accretion rate of $2.1 \times 10^{15}$ g/s, is given by
 
$T_{\rm eff} = 1.7 \times 10^4$~K ($<\dot{M}> / 10^{-10} )^{0.25} [ M_{\rm wd}/0.9])$.

Thus, the white dwarf should have $T_{\rm eff}= 14,272$ K, a factor of two smaller than from our modeling of ER UMa's FUV spectra in quiescence. For both the low inclination disk models and high inclination disk models, our estimates of the white dwarf surface temperature are a factor of two hotter than the average WD temperature of the white dwarfs in SU UMa systems below the period gap. 

The WD during quiescent accretion might be heated not only by compression due to the weight of the accreted gas but also by the boundary layer. If, as expected for low mass accretion rates, the boundary layer is optically thin, then it advects energy directly into the outer layers of the WD and especially in its equatorial region. The broad strong emission lines of C\,{\sc iii} (977 \& 1175 \AA ), O\,{\sc vi} (1131.9 \& 1137.6 \AA ), and also possibly of N\,{\sc iv} (921,46-924,91 \AA), S\,{\sc vi} (933.4 \& 944.5 \AA ) are all indicative of the presence a very hot component, namely the boundary layer.

\clearpage

\section*{Acknowlegements}  
We are grateful to an anonymous referee for helpful comments on ER UMa. 
P.G. is pleased to thank William (Bill) P. Blair at the Henry Augustus 
Rowland Department of Physics and Astronomy at the Johns Hopkins University,
Baltimore, MD, for is kind hospitality. 
This work is upported by the National Aeronautics and Space Administration 
(NASA) under grant number NNX17AF36G, issued through the Office of 
Astrophysics Data Analysis Program (ADAP) to Villanova University. 

{}  

\end{document}